\documentclass[12pt]{article}
\usepackage[english,german,french,polish]{babel}
\usepackage[T1]{fontenc}
\usepackage{amsfonts}

\begin{document}

\baselineskip 0.75cm
\topmargin -0.6in
\oddsidemargin -0.1in

\let\ni=\noindent

\renewcommand{\thefootnote}{\fnsymbol{footnote}}

\newcommand{\SM}{Standard Model }

\newcommand{\SMo}{Standard-Model }

\pagestyle {plain}

\setcounter{page}{1}

\pagestyle{empty}

~~~

\begin{flushright}
IFT-- 10/6
\end{flushright}

\vspace{0.3cm}

{\large\centerline{\bf How the tau-lepton mass can be understood{\footnote{Work supported in part by Polish MNiSzW scientific research grant N N202 103838 (2010--2012).}} }}

\vspace{0.5cm}

{\centerline {\sc Wojciech Kr\'{o}likowski}}

\vspace{0.3cm}

{\centerline {\it Institute of Theoretical Physics, University of Warsaw }}

{\centerline {\it Ho\.{z}a 69,~~PL--00--681 Warszawa, ~Poland}}

\vspace{1.2cm}

{\centerline{\bf Abstract}}

\vspace{0.3cm}

For some time the measured value of $\tau$-lepton mass continues to approach closer and closer a particular figure 1776.80 MeV predicted by an intrinsically composite model of three generations of leptons and quarks, introduced almost two decades ago. In the framework of this model, we recall briefly the argument standing behind our prediction.

\vspace{0.6cm}

\ni PACS numbers: 12.15.Ff , 12.90.+b .

\vspace{0.8cm}

\ni September 2010

\vfill\eject

~~~
\pagestyle {plain}

\setcounter{page}{1}

\vspace{0.2cm}

In 1992 we found for the charged-lepton masses the sum rule [1]

\vspace{-0.1cm} 

\begin{equation}
125 m_\tau = 6 \left(351m_\mu - 136 m_e\right) 
\end{equation}

\ni giving the prediction 

\vspace{0.2cm}

\begin{equation}
m_\tau  = 1776.80\;{\rm MeV} 
\end{equation}

\ni (or, more precisely, $m_\tau = 1776. 7964$ MeV), if the experimental values of $m_e$ and $m_\mu$ are used as the only input. It happens that since several years the measured value of the $\tau$-lepton mass $m_\tau$ converges on the particular figure (2), namely

\vspace{-0.2cm}

\begin{eqnarray} 
m^{(2004)}_{\tau} & = & 1776.99^{+0.29}_{-0.26}\;{\rm MeV} \;, \\
m^{(2006)}_{\tau} & = & 1776.99^{+0.29}_{-0.26}\;{\rm MeV} \;, \\
m^{(2008)}_{\tau} & = & 1776.84 \pm 0.17\;{\rm MeV} \,,\\
m^{(2010)}_{\tau} & = & 1776.82 \pm 0.16\;{\rm MeV} 
\end{eqnarray} 

\ni ([2], [3], [4], [5], respectively). In this note, for convenience of the Reader, we recall briefly the argument standing behind the prediction (2).

Our sum rule (1) follows strictly from the mass formula conjectured in 1992 for charged leptons [1],

\vspace{-0.2cm}

\begin{equation}
m_N  =  \rho_N \,\mu\! \left(\!N^2 + \frac{\varepsilon -1}{N^2}\! \right) \;\;\;\;(N = 1,3,5)
\end{equation}

\ni with three masses

\vspace{-0.2cm}

\begin{equation}
m_1  \equiv m_e\;,\;  m_3  \equiv m_\mu\;,\;  m_5  \equiv m_\tau,
\end{equation}

\ni where the specific constants

\vspace{-0.2cm}

\begin{equation} 
\rho_1 = \frac{1}{29} \;,\; \rho_3 = \frac{4}{29} \;,\; \rho_5 = \frac{24}{29} 
\end{equation}

\ni ($\sum_N\rho_N = 1$) may be called the generation-weighting factors, while $\mu$ and $\mu(\varepsilon - 1) $ stand for two mass-dimensional free parameters. The mass formula (7) can be rewritten explicitly as

\begin{eqnarray}
m_e & = & \frac{\mu}{29} \,\varepsilon  \,, \nonumber \\
m_\mu & = & \frac{4\mu}{29}\,\frac{80 +\varepsilon}{9}  \,, \nonumber \\
m_\tau & = & \frac{24\mu}{29}\,\frac{624 + \varepsilon}{25} \,, 
\end{eqnarray}

\ni implying then in a simple way the mass sum rule (1).

In fact, eliminating from Eqs. (10) two free parameters $\mu$ and $\varepsilon$ for $m_e$ and $m_\mu$, we obtain the sum rule (1) giving the prediction (2) for $m_\tau$,

\begin{equation}
m_\tau = \frac{6}{125} \left(351m_\mu - 136 m_e\right) = 1776.80\;{\rm MeV}, 
\end{equation}

\ni as well as determine two free parameters $\mu$ and $\varepsilon$, 

\begin{equation}
\mu = \frac{29 (9m_\mu - 4 m_e)}{320} = 85.9924\;{\rm MeV} \;\;,\;\;\; \varepsilon = \frac{320 m_e}{9 m_\mu - 4 m_e} = 0.172329 \,,
\end{equation}

\ni if the experimental values for $m_e$ and $m_\mu$ are used as the only input. We can see that the predicted $m_\tau$ is really close to the actual experimental estimate (6){\footnote{ In 1981, Koide proposed for 
charged-lepton masses another approach based on the neatly looking nonlinear equation symmetrical in $m_e$, $m_\mu$ and $m_\tau$ [6],
$$
m_e + m_\mu + m_\tau = \frac{2}{3}(\sqrt{m_e} + \sqrt{m_\mu} + \sqrt{m_\tau})^2,
$$
giving two solutions for $m_\tau$ in terms of experimental values of $m_e$ and $m_\mu$:
$$
m_\tau = \left[2(\sqrt{m_e} + \sqrt{m_\mu} )\pm \sqrt{3(m_e + m_\mu)+12\sqrt{m_e m_\mu}}\, \right]^2 = \left\{\begin{array}{r} 1776.97\,{\rm MeV} \\ 3.31735\,{\rm MeV}  \end{array} \right.\,,
$$ 
where $m_e$ = 0.5109989 MeV and $m_\mu$ = 105.65837 MeV. The first solution still agrees wonderfully with the central value of actual experimental estimate (6), though its small deviation from this experimental value is  larger than the tiny deviation of our prediction (11). The second solution gets no interpretation yet.}}.

It was argued that the conjectured charged-lepton mass formula (7) follows from an intrinsically composite model of three generations of fundamental fermions, leptons and quarks, numerated by the quantum number $N =1,3,5$ equal to the number of Dirac intrinsic partons involved. Such a model was introduced almost two decades ago [1,7] and is based on a generalized Dirac's square-root procedure leading to Dirac-type equations numerated by $N =1,3,5$. Here, only three values $N =1,3,5$ can appear, if a new intrinsic Pauli principle is applied to Dirac intrinsic partons within leptons and quarks. Such intrinsic partons are completely described by Dirac bispinor indices. Therefore, the latter are the formal subject of our Pauli principle.  

According to this argument, a linear combination of two terms: (i) an "$\!\,$intrinsic mutual and self-inter\-action"~of $N$ intrinsic partons of spin 1/2 (described by Dirac bispinor indices $ \alpha_1,\alpha_2,...,\alpha_N$) treated on equal footing, and (ii) a correction to the "$\!\,$intrinsic self-interaction"~of one of these intrinsic partons (described, say, by Dirac bispinor index $\alpha_1$), distinguished from all other $N-1$ partons by its presumed correlation with the \SMo label carried by leptons and quarks. Then, these other $N-1$ intrinsic partons, not correlated with the \SMo label, are mutually undistinguishable and thus, obey the Fermi statistics along with the intrinsic Pauli principle which antisymmetrizes $N-1$ Dirac bispinor indices $\alpha_2,...,\alpha_N$. Thus, $N = 1,3,5$ only, excluding larger odd $N$. 

In consequence, the first term is proportional to $N^2$, whereas the second proportional to $[N!/(N-1)!]^{-2} = N^{-2}$, as $N!/(N-1)!$ is the number of different states involved in the term (ii) (the inverse of this number squared gives the appearance probability of the correction (ii)). Therefore, with $m_N \propto \rho_N$, we get

\begin{equation}
m_N  =  \rho_N \left(a N^2 + b \frac{1}{N^2} \right) \;\;\;\;(N = 1,3,5) \,,
\end{equation}

\ni where $\rho_N$ are the generation-weighting factors which turn out to be given indeed as in Eqs (9). In such a way, with $a \equiv \mu$ and $b \equiv \mu(\varepsilon -1)$, the mass formula (7) follows from our intrinsically composite model [1,7].

We can see that the intrinsic Pauli principle, conjectured for all leptons and quarks, resricts $N$ to three values $N=1,3,5$ and so, the possible spectrum of leptons and quarks to three generations. This is our proposal for the mechanism of existence in Nature of exactly three lepton and quark generations [1,7].

In our intrinsically composite model there could appear also extra fundamental-fermion states, but including this time no distinguished intrinsic parton correlated with the \SMo label (and so, referred to the Dirac bispinor index $\!\alpha_1\!$). They would be identified with cold-dark-matter fermions (we called them "\,$\!$sterinos") which might form two generations of sterile Dirac fermions active in a hidden sector of the Universe [8].

\vfill\eject

~~~~
\vspace{0.5cm}

{\centerline{\bf References}}

\vspace{0.5cm}

{\everypar={\hangindent=0.6truecm}
\parindent=0pt\frenchspacing

{\everypar={\hangindent=0.6truecm}
\parindent=0pt\frenchspacing

~[1]~For a recent presentation {\it cf.} W.~Kr\'{o}likowski, {\it Acta Phys. Polon.}, {\bf B 41}, 649 (2010); {\it cf.} also {\it Acta Phys. Polon.}, {\bf B 38}, 3133 (2007); and references therein. 

\vspace{0.2cm}

~[2]~S.~Eidelman {\it et al.} ( Particle Data Group), {\it Phys. Lett.}, {\bf B 592}, 1 (2004).

\vspace{0.2cm}

~[3]~W.M.~Yao {\it et al.} ( Particle Data Group), {\it J. Phys}, {\bf G 33}, 1 (2006).

\vspace{0.2cm}

~[4]~C.~Amsler {\it et al.} (Particle Data Group), {\it Phys. Lett.} {\bf B 667}, 1 (2008).

\vspace{0.2cm}

~[5]~N.~Nakamura {\it et al.} (Particle Data Group), {\it J. Phys}, {\bf G 37}, 075021 (2010).

\vspace{0.2cm}

~[6]~For a more recent discussion {\it cf.} Y.~Koide, {\tt hep--ph/0506247}; and references therein.

\vspace{0.2cm}

~[7]~For a discussion {\it cf.} also W. Kr\'{o}likowski, {\tt hep--ph/0504256}; {\it Acta Phys. Polon.} {\bf B 37}, 2601 (2006); {\tt hep--ph/0604148}; and references therein.

\vspace{0.2cm}

~[8]~W. Kr\'{o}likowski, arXiv: 0811.3844 [{\tt hep-ph}]; {\it cf.} also {\it Acta Phys. Polon.}, {\bf B 38}, 3133 (2007); {\it Acta Phys. Polon.}, {\bf B 39}, 1881 (2008); {\it Acta Phys. Polon.}, {\bf B 40}, 111 (2009); arXiv: 0911.5614 [{\tt hep--ph}].

\vspace{0.2cm}

\vfill\eject

\end{document}